\documentclass[12pt]{article}
\usepackage{graphicx}
\usepackage[utf8]{inputenc}

\def\institute{On behalf of the ATLAS Collaboration \\ Physikalisches Institut\\
Universität Bonn, GERMANY}
\def\support{\footnote{Work Supported by European Research Council grant ERC-CoG-617185.}}

\def\Title#1{\begin{center} {\Large #1 } \end{center}}
\def\Author#1{\begin{center}{ \sc #1} \end{center}}
\def\Address#1{\begin{center}{ \it #1} \end{center}}

\newenvironment{Abstract}{\begin{quotation}  }{\end{quotation}}
\newenvironment{Presented}{\begin{quotation} \begin{center} 
             PRESENTED AT\end{center}\bigskip 
      \begin{center}\begin{large}}{\end{large}\end{center} \end{quotation}}
\def\Acknowledgements{\bigskip  \bigskip \begin{center} \begin{large}
             \bf ACKNOWLEDGEMENTS \end{large}\end{center}}

\def\beq{\begin{equation}}
\def\eeq#1{\label{#1}\end{equation}}
\def\eeqn{\end{equation}}

\def\beqa{\begin{eqnarray}}
\def\eeqa#1{\label{#1}\end{eqnarray}}
\def\eeqan{\end{eqnarray}}

\let\bar=\overbar

\def\Dslash{\not{\hbox{\kern-4pt $D$}}}
\def\dslash{\not{\hbox{\kern-2pt $\del$}}}

\def\msb{{\bar{\ssstyle M \kern -1pt S}}}

\begin{document}
\begin{titlepage}

\vfill
\Title{Measurement of the $t\bar{t}\gamma$ production cross section in
proton--proton collisions at $\sqrt{s} = 8$ Te\kern -0.1em V with the ATLAS
detector}
\vfill
\Author{ Mazuza Ghneimat\support}
\Address{\institute}
\vfill
\begin{Abstract}
The cross section of a top-quark pair produced in association
with a photon is measured in proton--proton collisions at a centre-of-mass
energy of $\sqrt{s} = 8$ Te\kern -0.1em V with $20.2$ fb$^{-1}$ of data
collected by the ATLAS detector at the Large Hadron Collider in 2012. The
measurement is performed by selecting events that contain a photon with
transverse momentum $p_\mathrm{T} > 15\;\mathrm{Ge\kern -0.1em V}$, an isolated
lepton with large transverse momentum, large missing transverse momentum, and
at least four jets, where at least one is identified as originating from a
$b$-quark. The production cross section is measured in a fiducial region close
to the selection requirements. It is found to be $139 \pm
7\,(\mathrm{stat.\!}\,) \pm 17\,(\mathrm{syst.\!}\,)\,\mathrm{fb}$, in good
agreement with the theoretical prediction at next-to-leading order of $151 \pm
24$ fb. In addition, differential cross sections in the fiducial region are
measured as a function of the transverse momentum and pseudorapidity of the
photon.

\end{Abstract}
\vfill
\begin{Presented}
$10^{th}$ International Workshop on Top Quark Physics\\
Braga, Portugal,  September 17--22, 2017
\end{Presented}
\vfill
\end{titlepage}
\def\thefootnote{\fnsymbol{footnote}}
\setcounter{footnote}{0}

\section{Introduction}
Measurements of top-quark properties play an important role in testing the
Standard Model (SM) and its possible extensions. Studies of the production and
dynamics of a top-quark pair in association with a photon ($t\bar{t}\gamma$) probe the
$t\gamma$ electroweak coupling. For instance, deviations in the spectrum of the transverse momentum $p_\mathrm{T}$ of
the photon from the SM prediction could point to new physics through anomalous
dipole moments of the top quark, as discussed in Refs.~\cite{Baur:2004uw,
Bouzas:2012av, Rontsch:2015una, Schulze:2016qas, Bylund:2016phk, Duan:2016qlc}.

Photons can originate not only from top quarks, but also from their decay
products, including the quarks and leptons from the decay of the $W$ bosons.
In addition, they can be radiated from incoming partons. The measurement of the $t\bar{t}\gamma$ production cross section is based on a data set recorded with the ATLAS detector~\cite{atlas} in 2012 at a centre-of-mass
energy of $\sqrt{s} = 8$ Te\kern -0.1em V and corresponding to an integrated
luminosity of $20.2$ fb$^{-1}$. The cross section is measured with a
maximum-likelihood fit using templates defined within a fiducial volume chosen
to be close to the selection requirements implemented in the analysis.  Only
final states with exactly one reconstructed lepton (electron or muon),
including those originating from $\tau$ lepton decays, are considered. These
final states are referred to as the single-lepton channel in the following. In
addition to the inclusive cross section, differential cross sections as a
function of $p_\mathrm{T}$ and the pseudorapidity $\eta$ of the
photon are measured for the same fiducial volume. The cross sections are
compared to the theoretical calculations at next-to-leading order
(NLO)~\cite{melnikov} in the strong interaction.

\section{Event selection}
\label{sec:sel}
A presence of exactly one lepton with $p_\mathrm{T} > 25\;\mathrm{Ge\kern -0.1em V}$, and at least four jets are required. The absolute pseudorapidity of the electron is requited to be less than 2.47, excluding the transition region between the barrel and the endcap of the electromagnetic calorimeter, 1.37 $< |\eta_{\mathrm{cl}}| < $ 1.52. In order to reduce several background contributions, but mainly the $W+$jets background, at least one jet is required to be tagged as a $b$-jet. For the muon channel, additional requirements on the missing transverse
momentum and on $m_{\mathrm{T}}^{W}$, the transverse mass of the $W$ boson
candidate~\cite{TOPQ-2010-01}, $E_\mathrm{T}^{\mathrm{miss}} > 20\;\mathrm{Ge\kern -0.1em V}$ and $E_\mathrm{T}^{\mathrm{miss}} + m_{\mathrm{T}}^{W} >
60\;\mathrm{Ge\kern -0.1em V}$, are imposed. For the electron channel the requirements are
tighter, $E_\mathrm{T}^{\mathrm{miss}} > 30\;\mathrm{Ge\kern -0.1em V}$  and $m_{\mathrm{T}}^{W} > 30\;\mathrm{Ge\kern -0.1em V}$, due to the larger
multijet background.

The $t\bar{t}\gamma$ candidates are selected by applying the above criteria and by
requiring exactly one photon with $p_\mathrm{T} > 25\;\mathrm{Ge\kern -0.1em V}$, and $|\eta| <$  2.37, excluding the transition region 1.37 $<  |\eta_{\mathrm{cl}}| < $ 1.52. Events with a jet within a cone of $\Delta R=0.5$
around the selected photon are rejected to remove photon radiation from quarks.
In order to enrich the sample with events in which a photon is radiated from a
top quark, the distance between the selected photon and lepton direction is
required to be larger than $\Delta R = 0.7$. For the electron channel, the invariant
mass of the electron and the photon has to be outside a $5\;\mathrm{Ge\kern -0.1em V}$ mass
window around the $Z$ boson mass (i.e. $m_{e\gamma} < 86\;\mathrm{Ge\kern -0.1em V}$ or
$m_{e\gamma} > 96\;\mathrm{Ge\kern -0.1em V}$) in order to suppress $Z$+jet events with one
electron misidentified as a photon. 

The fiducial region for this analysis is defined for Monte Carlo events at
particle level (before detector simulation), using the same event selections described above, except for $E_\mathrm{T}^{\mathrm{miss}}$, $m_{\mathrm{T}}^{W}$ and $m_{e\gamma}$ in order to obtain a common fiducial region for the electron channel and the muon channel.

\section{Background estimation}
\label{sec:bkg}
The main background contribution to the $t\bar{t}\gamma$ process comes from hadrons that are misidentified as prompt photons, or photons from hadron decays. This background is estimated from a control region in data with at least four jets. Events must have at least one photon candidate that fails to
satisfy at least one of the four photon identification criteria constructed
using shower-shape variables from the first layer of the electromagnetic calorimeter. 

Events with electrons misidentified as photons are the second largest
background contribution. This background contribution is the result of dileptonic processes where one electron is misidentified, dominantly from dileptonic $t\bar{t}$ events ($ee$ or $e\mu$), followed by smaller contribution form $Z$ boson decaying into an $e^+e^-$. The contribution of events with electrons misidentified as photons is estimated with a fully data-driven method. The probability for an electron to fake a photon is calculated
using two control regions, one enriched in $Z\rightarrow e^+e^- $ and one enriched in events
reconstructed as $Z\rightarrow e + \mathrm{fake}\: \gamma$ events. 


Processes with an associated prompt photon source constitute another source of background contribution. Those are estimated from $W \gamma + $jets, $Z \gamma + $jets, single top quark and diboson production, where all the contributions are obtained Monte Carlo simulations. Multijet production with an associated prompt photon is a source of background when one of the jets in the event is misidentified as a lepton. In order to estimate this background, a control sample is created that uses the same event selection as for the signal except that the lepton identification criteria are loosened. 

\section{Likelihood fit}
\begin{figure}[htbp]
\centering
\includegraphics[width=7cm,height=6cm]{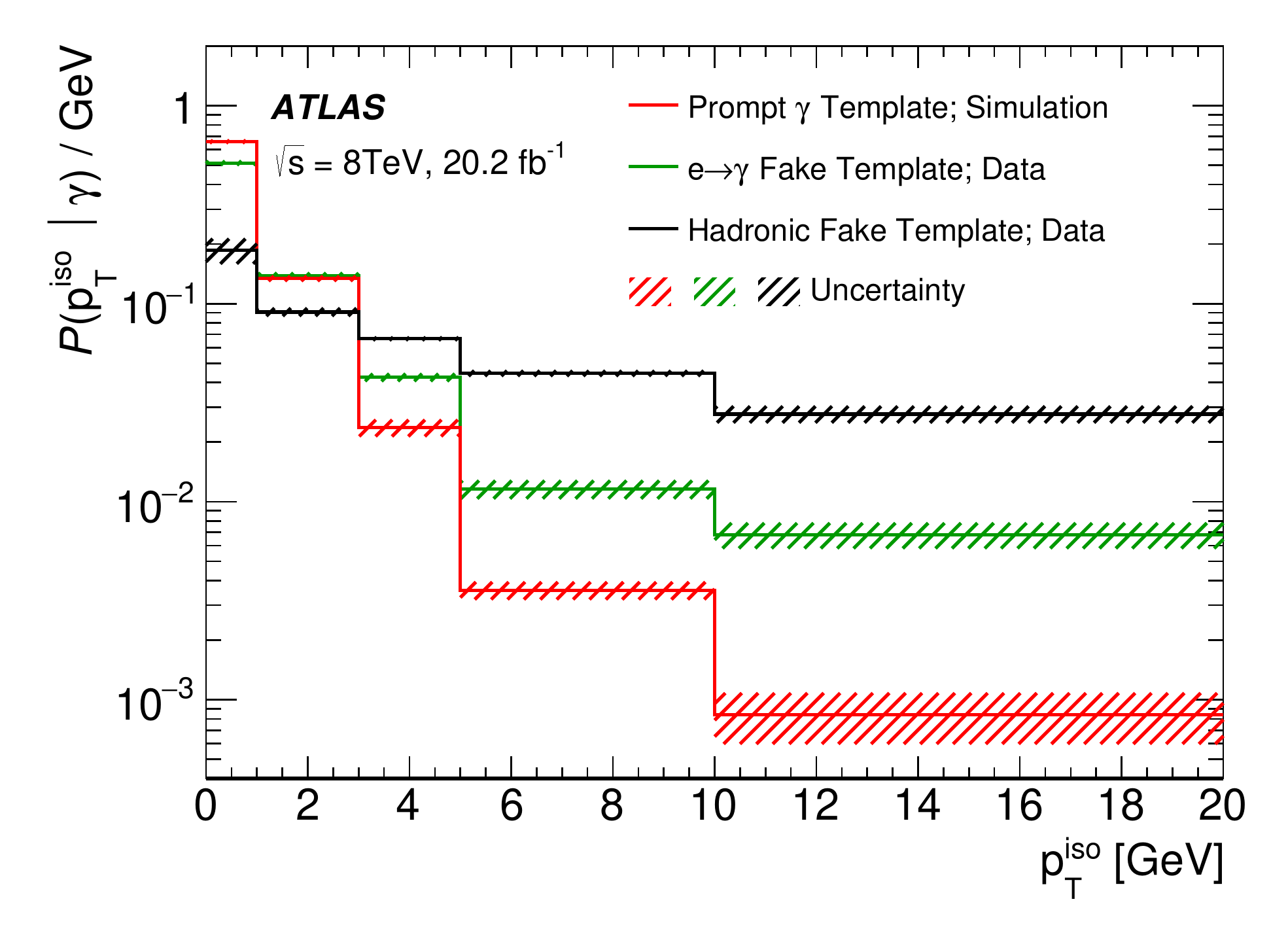}
\includegraphics[width=0.40\textwidth]{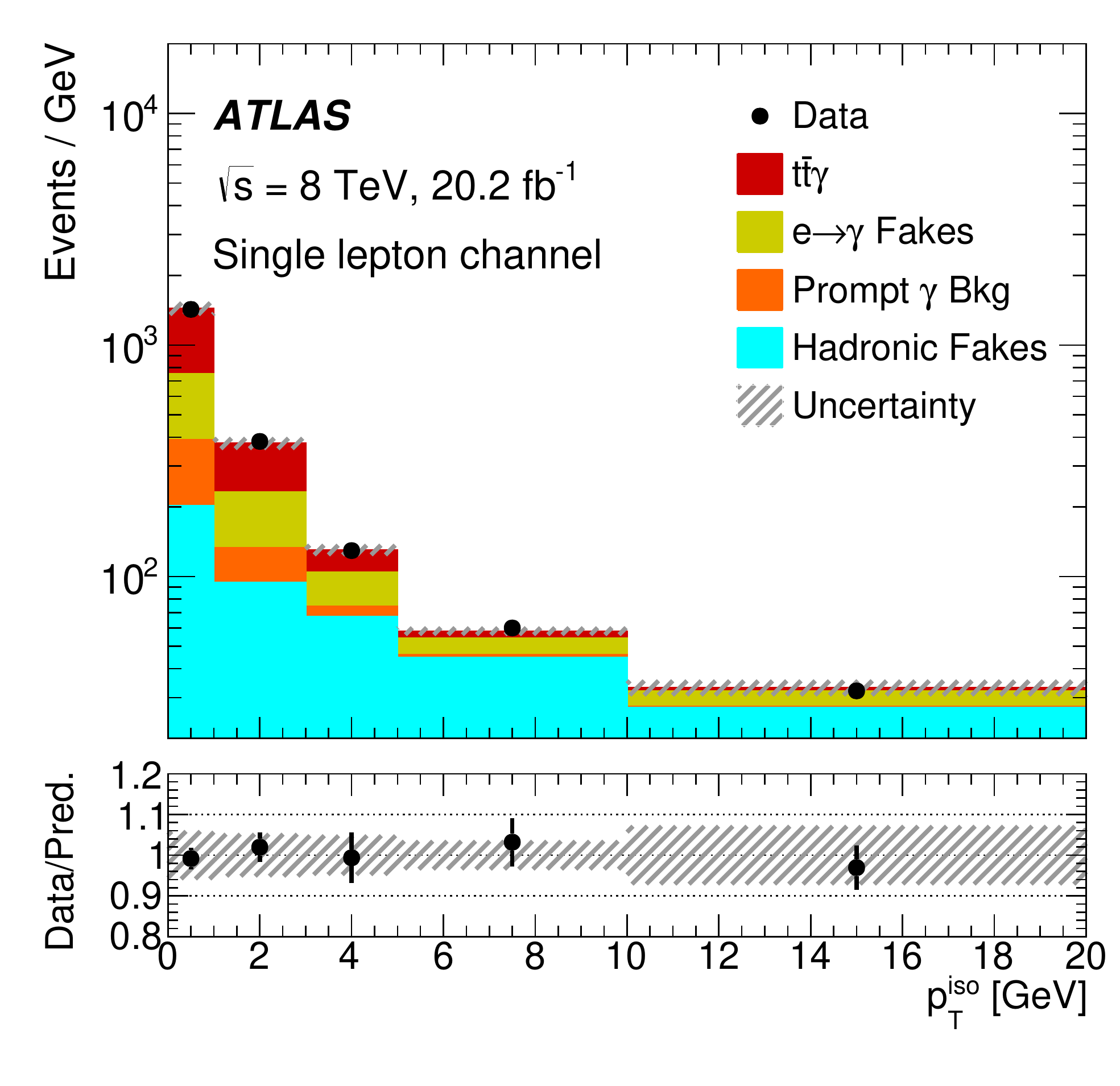}
\caption{The $p_\mathrm{T}^{\mathrm{iso}}$ templates for the inclusive cross-section measurement (left) for
prompt photons, hadronic fakes and electron fakes. The distributions are normalised to unity. The post-fit track isolation distribution for the inclusive cross-section measurement (right). The last bin includes
the overflow for both distribuitons and the uncertainty band includes all uncertainties~\cite{Aaboud:2017era}.}
\label{fig:Template}
\end{figure}

The extraction of the total and differential cross sections is
based on a likelihood fit using three templates (see Figure~\ref{fig:Template}), one for the prompt-photon
events, one for the hadronic-fake events and one for electrons misidentified as
photons. The normalisations of the first two templates are free parameters in
the likelihood fit, while for the third template the normalisation is fixed to
the data-driven estimate of the number of events with an electron misidentified
as a photon. The variable used for the templates is $p_\mathrm{T}^{\mathrm{iso}}$, the sum of the transverse momenta of all tracks within a cone with an opening angle around the photon of 0.2 rad.
This variable yields the best discrimination between signal and background and
has almost no dependence on the amount of pile-up~\cite{TOPQ-2012-07}.

All backgrounds, except for the hadronic fakes, are described by a Poisson probability distribution
with the mean values determined in Section~\ref{sec:bkg}. 





The inclusive and differential fiducial cross sections are related to the number of signal events by

\begin{equation}
  L \cdot \sigma_i \cdot C_i \cdot f_{i,j} = N_{i,j}^s,
\end{equation}

where $i$ corresponds to the bin in the differential distribution, $L$ is the integrated luminosity of the data sample, $\sigma_i$ is the fiducial cross section to be determined, $f_{i,j}$ is the fraction of events in bin $i$ falling into bin $j$ of $p_\mathrm{T}^{\mathrm{iso}}$, calculated from the signal
template, and $C_i$ is the ratio of the number of reconstructed events to the
number of generated events in the fiducial region in bin $i$. The ratio $C_i$ therefore corrects for
the event selection efficiency and the migration of events between the fiducial
and the non-fiducial region, including leptonic $\tau$ decays, or between different bins $i$, in the case of the differential cross-section measurement.

The largest systematic uncertainties on the measured cross section are due to the shape of the hadronic-fake template, and the estimation of events with one electron misidentified as a photon, each source results in an uncertainty of 6.3$\%$. 

The post-fit track isolation distribution for the inclusive measurement is shown in Figure~\ref{fig:Template}. 

\section{Results}

A total of $3072$ candidate events are observed in data. Using data selected in the single-lepton channel, the result of the inclusive measurement is:

\begin{equation}
 \sigma_{\mathrm{sl}}^{\mathrm{fid}} = 139  \pm  7~(\mathrm{stat}) \pm  17~(\mathrm{syst})~\mathrm{fb} = 139 \pm  18~\mathrm{fb},
 \nonumber
\end{equation}

which agrees with the NLO prediction of $151 \pm 24$ fb~\cite{melnikov}. 

\begin{figure}[htbp]
\centering
{
\includegraphics[width=0.45\textwidth]{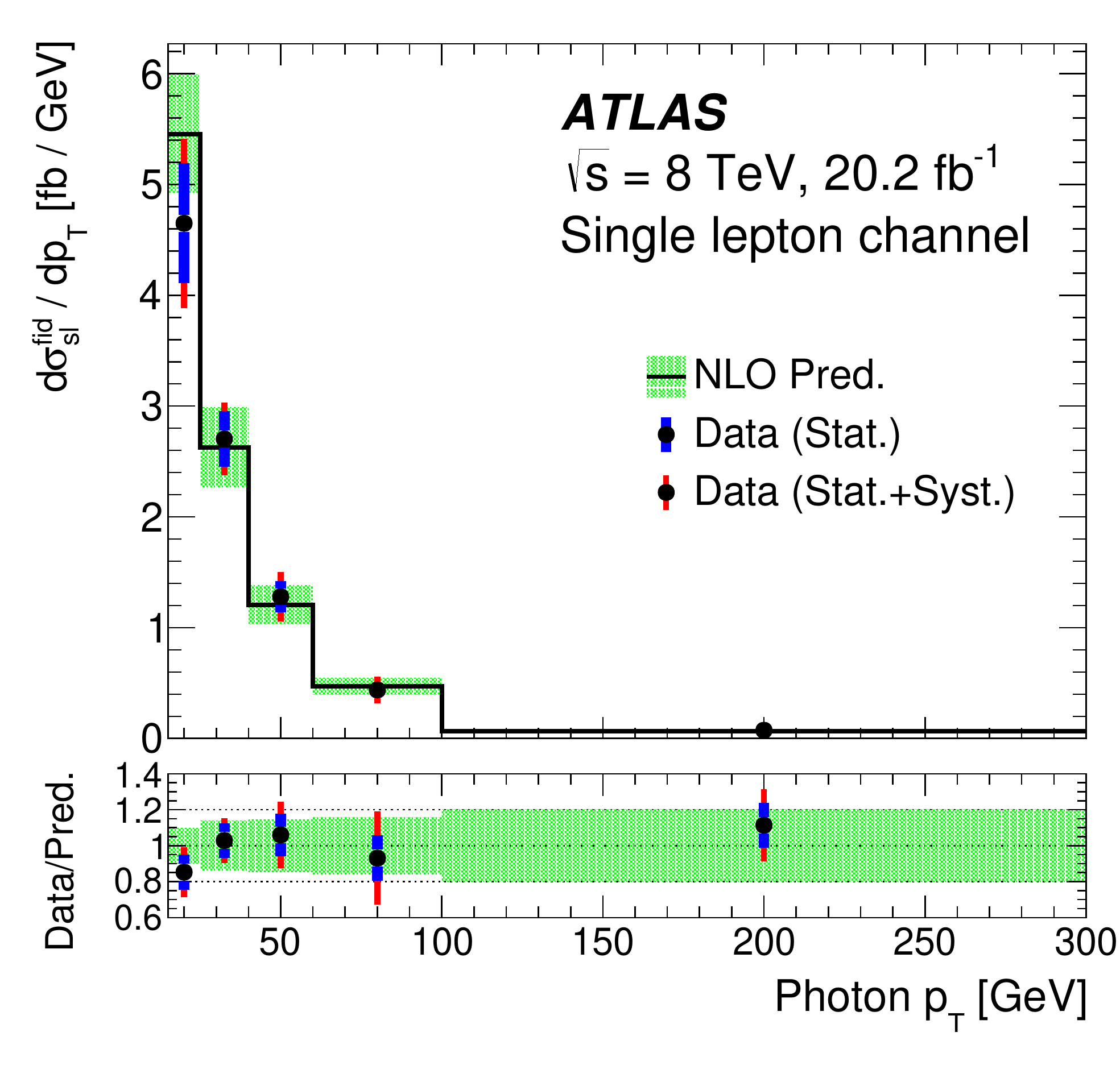}
\includegraphics[width=0.45\textwidth]{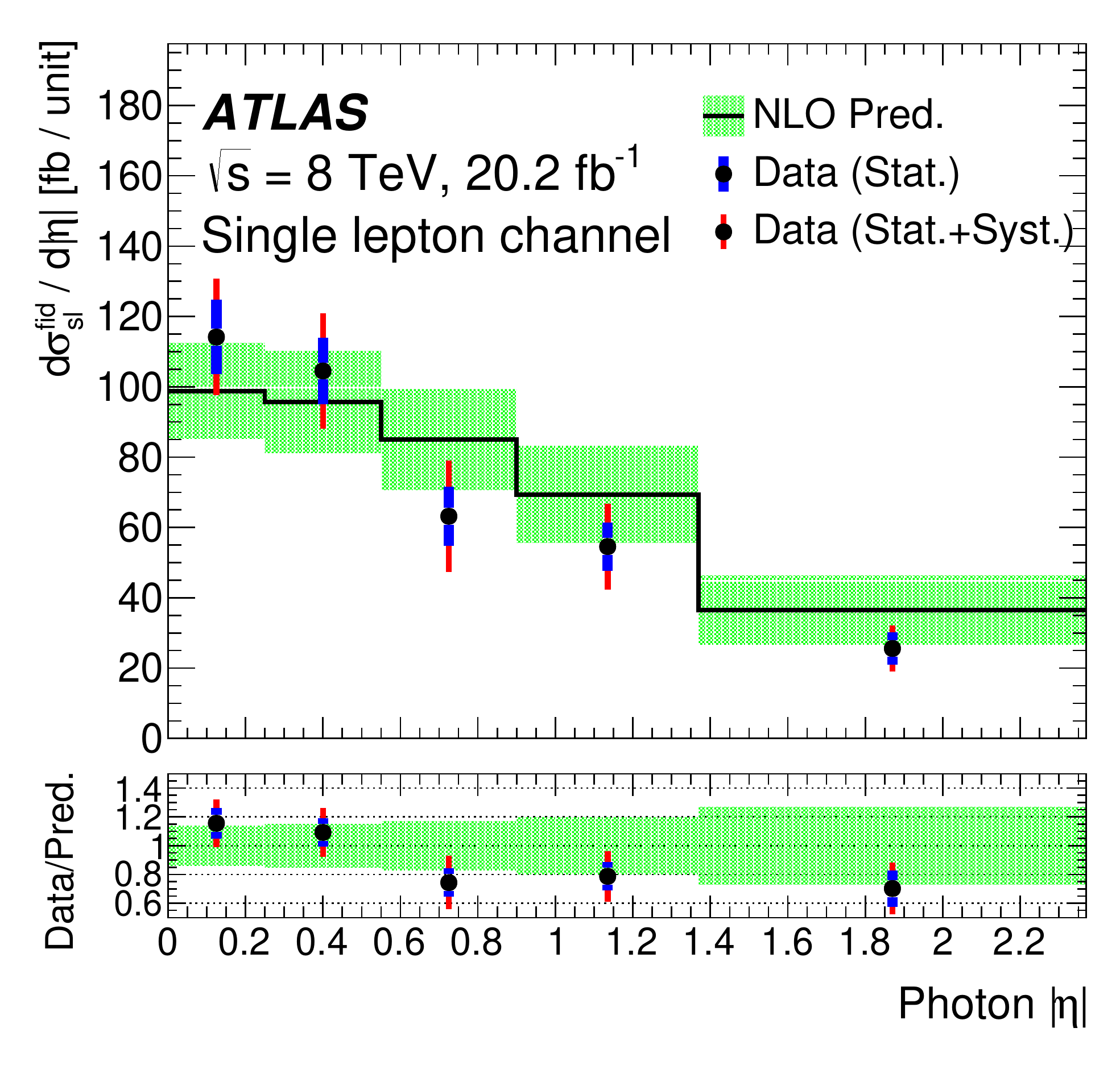}
}
\caption[]{Measured differential cross section in $p_\mathrm{T}$ (left)
and $|\eta|$ (right) and the corresponding theoretical
prediction~\cite{Aaboud:2017era}.} \label{fig:diff_xsec}
\end{figure}

The measured $p_\mathrm{T}$ and $\eta$ differential cross sections are shown and compared
with their corresponding theoretical predictions in Figure~\ref{fig:diff_xsec}.
Good agreement is observed between the measurement and predicted values.


\Acknowledgements
I thank the European Research Council for their grant and support to finish this work.

\end{document}